\RequirePackage{fix-cm}
\documentclass[twocolumn]{svjour3}          
\usepackage[a4paper, total={7.42in, 8in}]{geometry}

\smartqed  

\usepackage{multicol}
\usepackage{inputenc}
\usepackage{graphicx}
\usepackage{multicol,lipsum}
\usepackage{amsmath}
\usepackage{dcolumn} 
\usepackage{enumitem}
\usepackage{graphicx}
\usepackage{dcolumn}
\usepackage{array,multirow}
\usepackage{bm}
\usepackage{amsmath}
\usepackage{epstopdf}
\usepackage{amsfonts}
\usepackage{amssymb}
\usepackage{mathrsfs}
\usepackage{epsfig}
\usepackage{tabularx}
\usepackage{cite}
\usepackage[dvipsnames]{xcolor}
\RequirePackage[colorlinks,citecolor=blue,urlcolor=blue,linkcolor=blue]{hyperref}
\newcommand{\be}{\begin{equation}}
	\newcommand{\ee}{\end{equation}}
\newcommand{\beq}{\begin{eqnarray}}
	\newcommand{\eeq}{\end{eqnarray}}

\begin{document}

\title{Null geodesics in  five-dimensional Reissner-Nordstr\"om anti-de Sitter black holes}

\author{P. A. Gonz\'{a}lez  \and Marco Olivares \and
        Yerko V\'{a}squez \and J.R. Villanueva 
}

\institute{%
          P. A. Gonz\'{a}lez \at  
              Facultad de Ingenier\'{i}a y Ciencias,\\
              Universidad Diego Portales,\\ 
              Avenida Ej\'{e}rcito Libertador 441, Santiago, Chile.\\ 
              \email{\textcolor{blue}{\href{mailto:pablo.gonzalez@udp.cl}{pablo.gonzalez@udp.cl} }}
	\and Marco Olivares \at
              Facultad de Ingenier\'{i}a y Ciencias,\\ 
              Universidad Diego Portales,\\
              Avenida Ej\'{e}rcito Libertador 441, Santiago, Chile.\\
              \email{\textcolor{blue}{\href{mailto:marco.olivaresr@mail.udp.cl}{marco.olivaresr@mail.udp.cl} }}
    \and Yerko V\'{a}squez \at
              Departamento de F\'{\i}sica, Facultad de Ciencias,\\ Universidad de La Serena,\\ 
              Avenida Cisternas 1200, La Serena, Chile.\\
              \email{\textcolor{blue}{\href{mailto:yvasquez@userena.cl}{yvasquez@userena.cl} }}
    \and J.R. Villanueva \at
              Instituto de F\'isica y Astronom\'ia,\\ 
              Universidad de Valpara\'iso,\\
              Avenida Gran Breta\~na 1111, Valpara\'iso, Chile.\\
              \email{\textcolor{blue}{\href{mailto:jose.villanueva@uv.cl}{jose.villanueva@uv.cl} }}
}

\date{Received: date / Accepted: date}

\maketitle

\begin{abstract}
The study of the motion of photons around massive bodies is one of the most useful tools to know the geodesic structure associated with said gravitational source. In the present work, different possible paths projected in an invariant hyperplane are investigated, considering five-dimensional Reissner-Nordstr\"om anti-de Sitter black hole. 
Also, we study some observational test such as the bending of light and the Shapiro time delay effect.  Mainly, we found that the motion of photons follows the hippopede of Proclus geodesic,
which is a new type of trajectory of second kind, 
being the Lima\c{c}on of Pascal their analogue geodesic in four-dimensional Reissner-Nordstr\"om anti-de Sitter black hole.

\keywords{5-dimensional Black holes \and Geodesics \and Reissner-Nordstr\"om Anti-de Sitter}
\end{abstract}

\tableofcontents

\newpage

\section{Introduction}
\label{intro}

Extra-dimensional gravity theories have a long history that begins with an original idea propounded by Kaluza \& Klein \cite{Kaluza21,Klein26} as a way to unify the electromagnetic and gravitational fields, and nowadays finds a new realization within modern string theory \cite{Font:2005td, Chan:2000ms}. In spacetime dimensions $D \geq 4$, the spherically symmetric and static black hole solutions of general relativity in vacuum are known are Schwarzschild-Tangherlini black holes \cite{Tangherlini:1963bw}. Additionally, the most natural extension of general relativity to higher dimensions that generates field equations of second order is the Lovelock gravity. Remarkably, the action contains terms that appear as corrections to the Einstein-Hilbert action in the context of string theory. In five spacetime dimensions the Lovelock lagrangian is given by the Einstein-Hilbert term and the Gauss-Bonnet term, which is quadratic in the curvature and it is a topological invariant in four dimensions. An exact black hole solution to the field equations of Einstein-Gauss-Bonnet theory was found in \cite{Boulware:1985wk}. The geodesics of massive test particles in higher dimensional black hole spacetimes has been studied in Refs. \cite{Frolov:2003en,PhysRevLett.98.061102,Hackmann:2008tu,Gibbons:2011rh}, and it was shown that  a particular feature of Reissner-Nordstr\"om   spacetimes is that bound and escape orbits traverse through different universes, and the study of the motion of particles in five-dimensional spacetimes has been performed in Refs. \cite{Kovacs:1984qx, Seahra:2001bx, Guha:2010dd, Guha:2012ma, Kagramanova:2012hw, Chandler:2015aha,Gonzalez:2015qxc,Grunau:2017uzf, Kuniyal:2017dnu}.

The spacetime that we consider in this study is a generalization of Reissner-Nordstr\"om anti-de Sitter (RNAdS) black hole to five dimensions, that are interesting in the context of the AdS/CFT correspondence \cite{Maldacena:1997re, Witten:1998qj,Gubser:1998bc}.
Global five-dimensional Schwarzschild AdS solution was considered 
to describe a thermal plasma of finite extent expanding in a slightly anisotropic fashion \cite{Friess:2006kw}. 
Also, it was shown that four and five-dimensional charged black holes in AdS spacetime could be obtained by compactifications of the type IIB supergravity in eleven dimensions. The properties of Reissner-Nordstrom black hole in $d$-dimensional anti-de Sitter space-time has been studied in Refs. \cite{Chamblin:1999tk,Chamblin:1999hg}, and the null geodesic structure of four-dimensional RNAdS black holes was analytically investigated in Ref. \cite{Villanueva:2013zta}, where, concerning to the radial motion, it was shown that the photons arrive to the event horizon in a finite proper time, and infinite coordinate time, similar to the Schwarzschild case. Also, concerning to the 
 angular motion of photons it was shown that there are five different kinds of motion for trapped photons depending on the impact parameter of the orbits, that corresponds to 
 orbits where the photon arrives from infinity and falls into the event horizon, photons moving along the critical orbits that represent trajectories that come from infinity and fall asymptotically into a circle, photons falling from infinity arriving to some minimal distance and then going back to the infinity again, photon orbits described by Pascal Lima\c{c}on, which is an exclusive solution of black hole with the cosmological constant but it does not depend of the value of the cosmological constant, and finally confined orbits for the photons.

 The aim of this work is to study the null geodesics in a five-dimensional charged black hole, and to see if it is possible to find orbits for the motion of photons different to the previously mentioned for a RNAdS spacetime.  Here, we will find the null structure geodesic analytically, and interestingly enough we find a new kind of orbit called {\it{"Hippopede geodesics"}}, that under our knowledge is the first time that has been reported in the literature.

The five-dimensional spacetime considered allows us to study the role of extra dimensions, for instance, five-dimensional Myers-Perry black hole spacetime was studied in Ref. \cite{Kuniyal:2017dnu}, where the metric describes a spacetime with two spin parameters, and it was found that circular orbit geodesics are allowed, as well as, the deflection angle and the strong deflection limit coefficients differs from four-dimensional Kerr black hole spacetime due to the presence of two spin parameters in higher dimension. Other spacetime studied, corresponds to a geometry described by a spherically symmetric four-dimensional solution embedded in a five-dimensional space known as a brane-based spherically symmetric solution analyzed in Ref. \cite{Gonzalez:2015qxc}, where the authors found that the extra dimension contributes to the existence of bounded orbits for the photons, such as planetary and circular stable orbits. The spacetime considered in this work could be compared with four-dimensional RNAdS  black holes, 
for the five-dimensional spacetime  there is not an additional parameters apart of the dimension added, but the event horizon is not the same due to the change in the lapse function,
which could explain the differences between four and five-dimensional spacetimes. However, as we will see, the effect of additional dimensions could be the existence of the hippopede of Proclus geodesic found here, versus its analogue geodesic in four-dimensional RNAdS black hole, i.e, the Lima\c{c}on of Pascal \cite{Villanueva:2013zta}, both trajectories of second kind.

 It is worth mentioning that the same spacetime was considered in Ref. \cite{Guha:2010dd}, where the null geodesics were studied from the point of view of the effective potential formalism and the dynamical systems approach. The radial and circular trajectories were investigated, and it was found that photons will trace out circular trajectories for only two distinct values of specific radius of the orbits. The dynamical systems analysis was applied to determine the nature of trajectories and the fixed points, and it was shown that the null geodesics have a unique fixed point and these orbits are terminating orbits. Also, the thermodynamics and the stability of the spacetime under consideration were studied via a thermodynamic point of view, and it was found special conditions on black hole mass and black hole charge where the black hole is in stable phase \cite{Saadat:2012zza}.

The paper is organized as follows. In section \ref{GLBHS} we give a brief review of the spacetime considered. Then, in Sec. \ref{nullgeod}, we establish the null structure and we perform some test as the bending of light and the Shapiro time delay effect. Finally, we conclude in Sec. \ref{conclusion}.

\section{Five-dimensional Reissner-Nordstr\"om anti-de Sitter black holes}
\label{GLBHS}

Schwarzschild and Reissner-Nordstr\"om black hole solutions in $d$ spacetime dimensions were presented by Tangherlini \cite{Tangherlini:1963bw}. 
The five-dimensional RNAdS
black holes are solutions of the equations of motion that arise from the action \cite{Chamblin:1999hg}
\begin{equation}
    S=-\frac{1}{16\pi G_5}\int d^5x\sqrt{-g}(R-2\Lambda-F^2)\,,
\end{equation}
where $G_5$ is the Newton gravitational constant in five-dimensional spacetime, $R$ is the Ricci scalar, $F^2$ represents the electromagnetic Lagrangian, and $\Lambda=-6/\ell^2$, is the cosmological constant where $\ell$ is the radius of AdS$_5$ space. The static and spherically symmetric metric that solves the field equation derived from the above action is given by
\begin{equation}\label{metr}
{\rm d}s^2=-f(r)\,{\rm d}t^2+\frac{1}{f(r)}{\rm d}r^2+r^2\,{\rm d}\Omega_{3}^2\,,
\end{equation} where $f(r)$ for a $(n+1)$-dimensional RNAdS spacetime is the lapse function given by 

\begin{equation}
    f(r)=1-\frac{m}{r^{n-2}}+\frac{q^2}{r^{2n-4}}+\frac{r^2}{\ell^2},
\end{equation}
where $m$, and $q$ are arbitrary constant, and ${\rm d}\Omega_{3}^2={\rm d}\theta^2+\sin^2 \theta \,{\rm d}\phi^2+\sin^2 \theta \, \sin^2 \phi\, {\rm d}\psi^2 $ is the metric of the unit 3-sphere. Also, $m$ is related to the ADM mass $\mathcal{M}$ of the spacetime through
\begin{equation}
    \mathcal{M}=\frac{(n-1)\omega_{n-1}}{16\pi G}m\,,
\end{equation}
where $\omega_{n-1}$ is the volume of the unit $(n-1)$-sphere. The parameter $q$ yields the charge
\begin{equation}
    \mathcal{Q}=\sqrt{2(n-1)(n-2)}\left(\frac{\omega_{n-1}}{8 \pi G}\right)q\,.
\end{equation}
In this work, we consider $n=4$, $m\rightarrow (2M)^2$, and $q^2\rightarrow Q^4$, so the metric is 
\begin{equation}
\label{lapse} 
f(r)=1-\left( \frac{2M}{r}\right)^2+\left( \frac{Q^2}{r^2}\right)^2+\left( \frac{r}{\ell} \right)^2\,,
\end{equation}
thereby, $M$ and $Q$ are related to the total mass $\mathcal{M}$ and the charge $\mathcal{Q}$ of the spacetime via the relations
\begin{equation}
    (2M)^2=\frac{16\pi G \mathcal{M}}{(n-1)\omega_{n-1}}\,,\, Q^2=\frac{8 \pi G \mathcal{Q}}{\sqrt{2(n-1)(n-2)}\omega_{n-1}}\,.
\end{equation}

\begin{figure}[!h]
	\begin{center}
		\includegraphics[width=80mm]{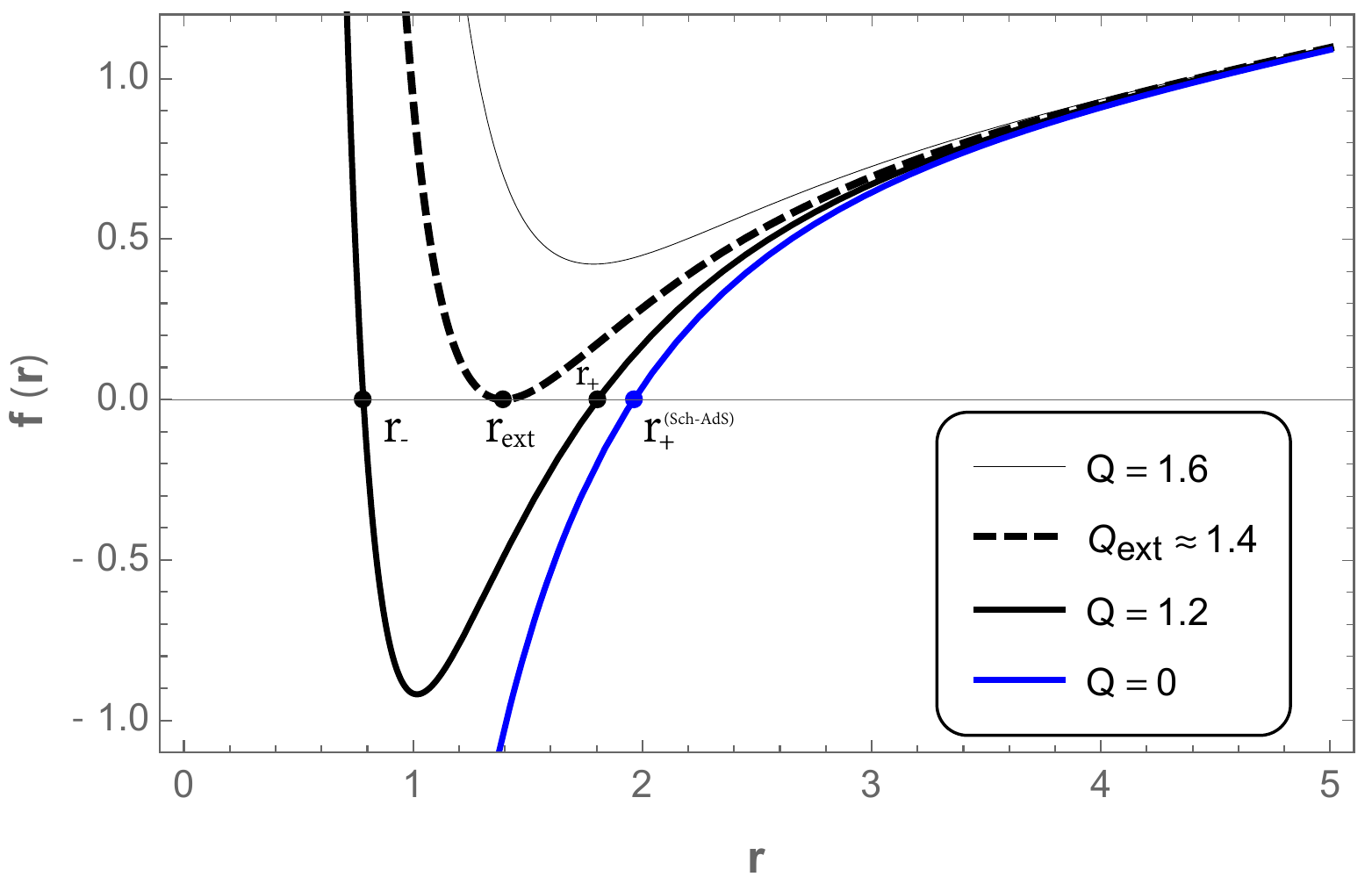}
	\end{center}
	\caption{The behaviour of the metric function $f(r)$, with $M=1$, $\ell=10$, for different values of $Q$. }
\label{lapsus}
\end{figure}

This spacetime allows two horizons (the event horizon $r_+$, and the Cauchy horizon $r_{-}$), which are obtained from the  equation $f(r)=0$, or
\begin{equation}
P(r) \equiv r^6+\ell^2\,r^4-4 M^2 \ell^2\,r^2+\ell^2 Q^4=0\, .
\end{equation}
Now, with the change of variable $x=r^2-\ell^2 /3$, we obtain
$P(x)=x^3-\alpha x+ \beta$,
where
\begin{equation}
\alpha=\ell^2\left(4M^2+ \frac{\ell^2}{3}\right) 
, \,\,\, \beta=\ell^2\left( Q^4+\frac{4M^2\ell^2}{3}+\frac{2\,\ell^4}{27}\right)\,,
\end{equation}
and the event and Cauchy horizons are given, respectively, by
\begin{eqnarray}\label{sol1}
r_{+}&=&\sqrt{\xi_0\cos \xi_1-{\ell^2\over3}}, \\\label{sol2} 
r_{-}&=&\sqrt{{\xi_0\over2}\left(\sqrt{3}\sin \xi_1- \cos \xi_1\right) -{\ell^2\over3}}\,,
\end{eqnarray}
where $\xi_0=2 \sqrt{\alpha/3}$ and $\xi_1=\frac{1}{3} \arccos \left(- \frac{3 \beta}{2 }\sqrt{\frac{3}{\alpha^{3}}} \right)$. Also, the extremal black hole is characterized by the degenerate horizon $r_{ext} =r_{+ }=r_{-}$, which is  obtained when: 
\begin{equation}
\label{equ}
4\ell^4\left(4M^2-Q^2 \right) +8\ell^2M^2\left(32M^4-9Q^4 \right)-27Q^8 =0\,.     
\end{equation}
In Fig. \ref{lapsus}, we plot curves for different values of $Q$ that show the behaviour of the lapse function against $r$, we can observe that when the charge of the black hole $Q$ increases we have a transition from a black hole to a naked singularity, passing by the extremal case.  

  Note that when $Q=0$, the lapse function reduces to the five-dimensional Schwarzschild anti-de Sitter black hole, and the spacetime allows one horizon (the event horizon $r_+$) given by $$r_+=\ell\,\sinh \left[{1\over 2} \sinh^{-1}\left({4M\over \ell} \right) \right].$$
 In Fig. \ref{lapsus}, the blue line corresponds to the case $Q=0$, we can observe that, for the same values of $\Lambda$ and $M$, the event horizon is greater for an uncharged than for charged black hole.

\section{The null structure}\label{nullgeod}
In order to obtain a description of the allowed motion in the exterior spacetime of the black hole, we use the standard Lagrangian formalism \cite{Chandrasekhar:579245,Cruz:2004ts,Villanueva:2018kem}, so that, the corresponding Lagrangian associated with the line element (\ref{metr}) reads

\begin{equation}\label{Lag1}
\mathcal{L}=-\frac{f(r)\, \dot{t}^2}{2}+ \frac{\dot{r}^2}{2f(r)}+\frac{r^2}{2} \mathcal{L}_{\Omega}\,,  \end{equation}
where $\mathcal{L}_{\Omega}$ is the {\it angular Lagrangian}:
\begin{equation}\label{Lag2} 
\mathcal{L}_{\Omega}= \dot{\theta}^2+ \sin^2 \theta\,\dot{\phi}+\sin^2 \theta\,\sin^2 \phi\, \dot{\psi}^2\,,
\end{equation}
and the dot indicates differentiation with respect to an affine parameter $\lambda$ along the geodesic.  Since the Lagrangian (\ref{Lag1}) does not depend on the coordinates ($t,\psi$), they are {\it cyclic coordinates} and, therefore, the corresponding conjugate momenta $\pi_{q} = \partial \mathcal{L}/\partial \dot{q}$ are conserved. Explicitly, we have 
\begin{eqnarray}\label{pp1}
\pi_{t}&=& -f(r)\,, \dot{t}\equiv -E\,, \\\label{angcons} 
 \pi_{\psi}&=&r^2\sin^2 \theta \sin^2 \phi\, \dot{\psi}=L\,,
\end{eqnarray}where $E$ is a positive constant that depicts the temporal invariance of the Lagrangian, which cannot be associated with  energy because the spacetime defined by the line element (\ref{metr}) is not asymptotically flat, whereas the constant $L$ stands the conservation of angular momentum, under which it is established that the motion is performed in an invariant hyperplane. Here, we claim to study the motion in the invariant hyperplane $\theta=\phi = \pi/2$, so $\dot\theta =\dot\phi=0$ and, from Eq. (\ref{angcons}), 
\begin{equation} \dot{\psi}=\frac{L}{r^2}\,.
\label{eq1} 
\end{equation}
Therefore, using the fact that $\mathcal{L}=0$ for photons together with Eqs. (\ref{pp1}) and (\ref{angcons}), we obtain the following equations of motion 

\begin{eqnarray}
&&\left(\frac{{\rm d}r}{{\rm d} \lambda}\right)^{2}= E^2-V^2(r)\,,\\
\label{w.12}
&&\left(\frac{{\rm d} r}{{\rm d} t}\right)^{2}= {f^{\,2}(r)\over E^2}\left[E^2-V^2(r)\right]\,,\\
\label{w.13}
&&\left(\frac{{\rm d} r}{{\rm d} \psi}\right)^{2}=  {r^{\,4}\over L^2}\left[E^2-V^2(r)\right]\,,
\label{w.14}
\end{eqnarray}
where the effective potential $V^2(r)$ is defined by
\begin{equation}\label{tl8}
V^2(r)\equiv L^2\,\frac{f(r)}{r^2}=\frac{L^{2}}{\ell^{2}}+\frac{L^{2}}{r^{2}}-\frac{4M^2 L^2}{r^4}+\frac{Q^{4} L^2}{r^{6}}\,.
\end{equation}
The effective potential for five-dimensional\\ Schwarzschild anti-de Sitter black hole is obtained by setting $Q=0$ in the above equation.

\subsection{Radial motion}

\begin{figure}[!h]
	\begin{center}
		\includegraphics[width=90mm]{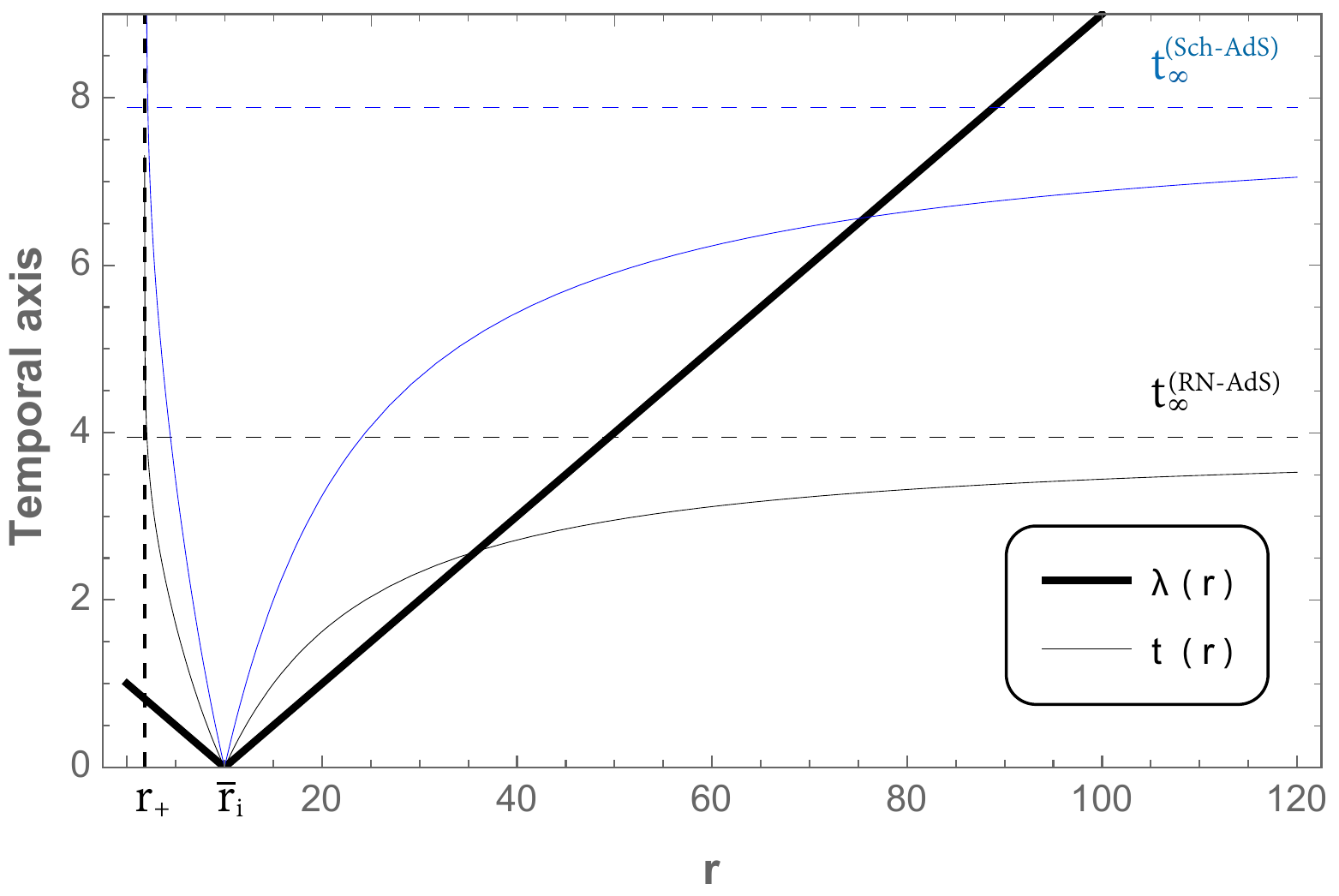}
	\end{center}
	\caption{Plot of the radial motion of massless particles.
		Particles moving to the event horizon, $r_+$, cross it in a  finite affine parameter, but an external observer will see that photons
		take an infinite (coordinate) time to do it.
		Here we have used the values $E=100$, $\ell=10$, and $\bar{r}_i=10$. Black-thin line for $Q=1.2$, and	$r_+\approx 1.80$. Blue line for $Q=0$, and  $r_+\approx 1.96$ }
	\label{f2}
\end{figure}

For the radial motion the condition $L=0$ holds, which immediately yields to a vanishing effective potential, $V^2=0$. Consequently, the equations governing this kind of motion are
\begin{equation}
\frac{dr}{d\lambda}=\pm E\,,
\label{mr.1}
\end{equation}
and
\begin{equation}
\frac{dr}{dt}=\pm f(r)\,,
\label{mr.2}
\end{equation}
where the sign $+$ ($-$) corresponds to massless particles  moving
toward the spatial infinite (event horizon).
Assuming that photons are placed at $r=\bar{r}_i$
when $t=\lambda=0$, a straightforward integration of Eq. (\ref{mr.1}) yields
\begin{equation}
\lambda(r)=\pm\frac{r-\bar{r}_i}{E}\,,
\label{mr.3}
\end{equation}
which is plotted in Fig. \ref{f2}. We observed that respect to the affine parameter the photons arrive at the horizon in a finite affine parameter, and 
when the photons move in the opposite direction, they require an infinity affine parameter  to  arrive to infinity, which does not depend with the charge of the black hole.

This behaviour is essentially the same as that reported for the 4-dimensional counterpart \cite{Villanueva:2013zta}. 
On the other hand, Eq. (\ref{mr.2}) can be arranged and then  integrated leading to the following expression
\begin{equation}
t(r)=\pm\,\ell^2\,\sum_{j=1}^{3}\delta_j\,t_j(r)\,,
\label{mr.5}
\end{equation}
where the functions $t_j(r)$ are given explicitly by
\begin{eqnarray}
\label{inv1} t_1(r)&=&\ln \left|\frac{\bar{r}_i+r_{+}}{\bar{r}_{i}-r_{+}}\frac{r-r_{+}}{r+r_{+}}\right|\,,\\  t_2(r)&=&\ln \left|\frac{\bar{r}_i+r_{-}}{\bar{r}_{i}-r_{-}}\frac{r-r_{-}}{r+r_{-}}\right|\,,\\
t_3(r)&=&\tan^{-1}(r/R)-\tan^{-1}(\bar{r}_i/R)\,,
\end{eqnarray}
with the corresponding constants,
\begin{eqnarray}
\label{inv2} \delta_1&=&{r_+^3\over 2(r_+^2-r_-^2)(r_+^2+R^2)}\,,\\  \delta_2&=&{-r_-^3\over 2(r_+^2-r_-^2)(r_-^2+R^2)}\,,\\
\delta_3&=&{R^3\over (r_+^2+R^2)(r_-^2+R^2)}\,,\\
R&=&\sqrt{\ell^2+r_+^2+r_-^2}\,.
\end{eqnarray}
Thus, an observer located at $\bar{r}_i$ will measure an infinite time for the photon to reach the event horizon, which also occurs in 3+1 dimensions.
Nevertheless, when the test particles move in the opposite direction, they require a finite coordinate time for arrive  to infinity given by the relation
$$t_{\infty}=\lim_{r\rightarrow \infty }t\left( r\right)$$
or, explicitly (with $\tilde{t}_{\infty}\equiv t_{\infty}/\ell^2-\delta_3 \pi/2$)
\begin{equation}\label{n6}
\tilde{t}_{\infty}= \delta_1\ln \left|\frac{\bar{r}_i+r_{+}}{\bar{r}_{i}-r_{+}}\right| +\delta_2\ln\left|\frac{\bar{r}_i+r_{-}}{\bar{r}_{i}-r_{-}}\right|-\delta_3 \arctan\left( {\bar{r}_i\over R}\right).
\end{equation}%
All previously described by Eqs. (\ref{mr.3}) and (\ref{mr.5}) is shown in Fig. \ref{f2}. It is interesting to note that the behaviour given in (\ref{n6}) also appears in Lifshitz space-times \cite{Olivares:2013uha,Cruz:2013ufa}, where it was argued that this corresponds to a general behavior of these manifolds \cite{Villanueva:2013gra}, and also occurs in the three-dimensional rotating Hořava-AdS black hole \cite{Gonzalez:2019xfr}.

On the other hand, for $Q=0$ Eq. (\ref{mr.3}) is valid. However, the solution for the coordinate time is given by 
\begin{equation}
t(r)=\pm\,\ell^2\,\sum_{j=1}^{2}\zeta_j\,t_j(r)\,,
\end{equation}
\begin{eqnarray}
t_1(r)&=&\ln \left|\frac{\bar{r}_i+r_{+}}{\bar{r}_{i}-r_{+}}\frac{r-r_{+}}{r+r_{+}}\right|\,,\\
t_2(r)&=&\tan^{-1}\left[ {r\over (r_+^2+\ell^2)^{1\over 2 }} \right] -\tan^{-1}\left[ {\bar{r}_i\over (r_+^2+\ell^2)^{1\over 2}} \right]\,, 
\end{eqnarray}
and
\begin{eqnarray}
\zeta_1&=&{r_+\over 2(2r_+^2+\ell^2)},\\  \zeta_2&=&{(r_+^2+\ell^2)^{1\over 2 }\over 2r_+^2+\ell^2}\,.
\end{eqnarray}
Also, in the asymptotic region, $r\rightarrow \infty$, the time to arrive at infinity reduces to 
\begin{equation}
t_{\infty}=\ell^2 \zeta_1\ln \left|\frac{\bar{r}_i+r_{+}}{\bar{r}_{i}-r_{+}}\right| +\ell^2\zeta_2 \left[  {\pi \over 2}-\tan^{-1}\left( {\bar{r}_i\over R}\right) \right]\,.
\end{equation}

It is possible to observe in Fig \ref{f2} that an observer located at $\bar{r}_i$ will measure an infinite coordinate time for the photon to reach the event horizon, and it does not depend on the charge of the black hole. However, when the photons move in the opposite direction, they require a finite coordinate time to  arrive  to infinity, which decreases with the charge of the black hole.

\subsection{Angular motion}\label{angmot}
Now we study the motion with $L\neq0$, so we put our attention in  Eq. (\ref{w.14}), which, after using (\ref{tl8}), is conveniently written as 
\begin{eqnarray}\nonumber
 \left(r \frac{{\rm d}r}{{\rm d}\psi}\right)^2&=&\left(\frac{1}{b^2}-\frac{1}{\ell^2}\right)\,r^6-r^4+4M^2\,r^2-Q^4\\ \label{eqmot}
 &=&\frac{r^6}{\mathcal{B}^2}-r^4+4M^2\,r^2-Q^4\,,
\end{eqnarray}where $b\equiv L/E$ is the impact parameter and $\mathcal{B}$ is the {\it anomalous impact parameter}, which is a typical quantity of the Anti-de Sitter spacetimes \cite{Cruz:2004ts}.

\begin{figure}[!h]
	\begin{center}
		\includegraphics[width=90mm]{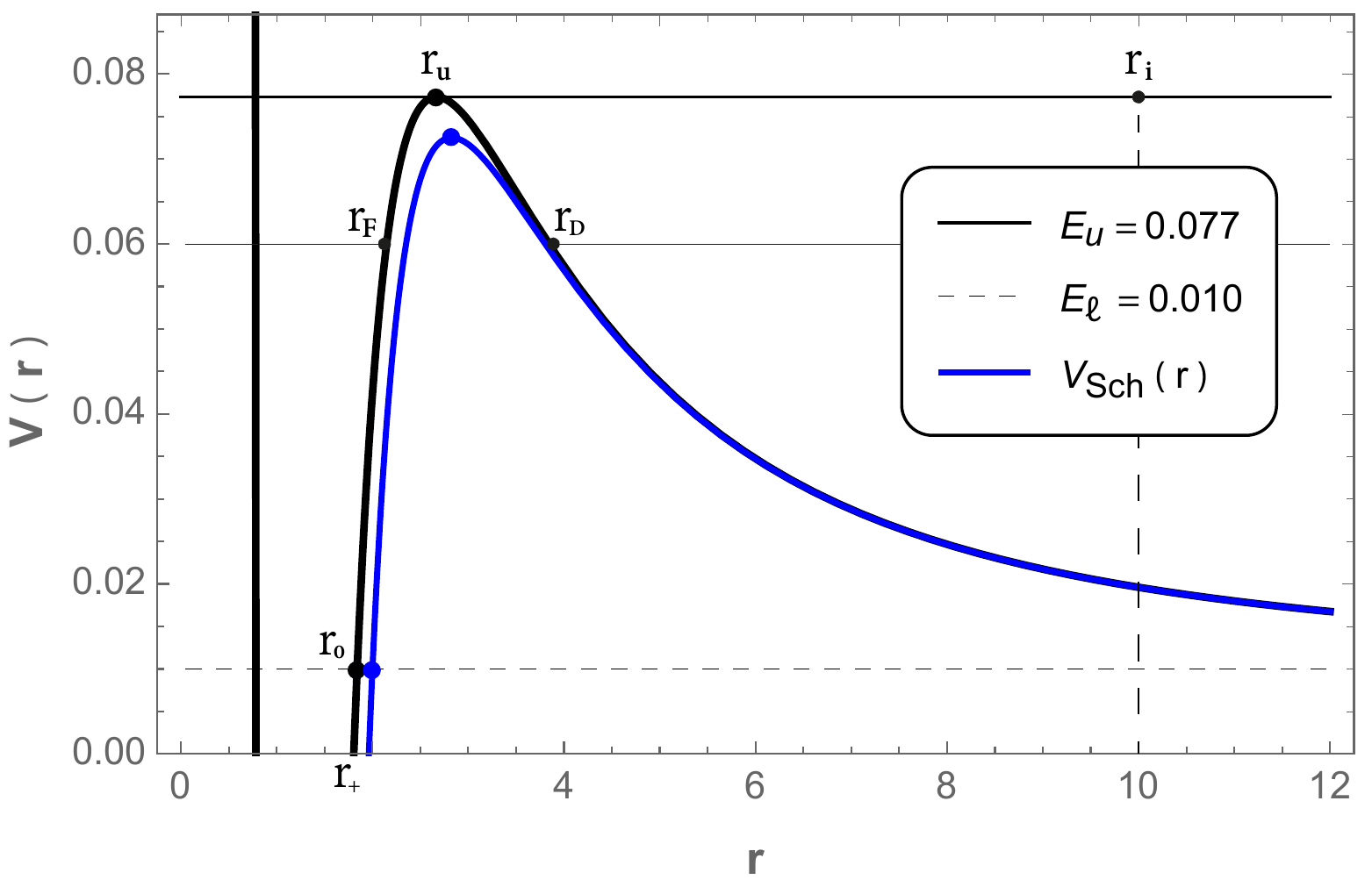}
	\end{center}
	\caption{Plot of the effective potential of photons.
		Here we have used the values $L=1$, $M=1$, $\ell=10$, $r_i=10$, $Q=1.2$ (black line), and $Q=0$ (blue line).}
	\label{f3}
\end{figure}

As a first approach, it is necessary to perform a qualitative analysis of the effective potential. So, we can observe in Fig. \ref{f3} the existence of a maximum potential
placed at 
\begin{equation}
    \label{max1}r_u=2\sqrt{2}M\cos\left[ {1\over 2}\sin^{-1}\left({\sqrt{3}Q^2\over 4M^2} \right) \right].
\end{equation}
So, $E_u$ is given by $V(r_u)$, and it corresponds to the energy of the photons for which the  potential is maximum. Also, it is possible to define $E_{\ell}$ given by $V(r\rightarrow \infty)=L/\ell$, and $b_{\ell}=L/E_{\ell}=\ell$ as its impact parameter. Thus, for orbits of first kind, the parameter $b_{\ell}$ is not allowed for photons, and  the deflection of the light is allowed for $E_{\ell}<E<E_u$ ($b_u<b<\ell$), see Fig. \ref{f3}; thereby, the radius of AdS$_5$ space $\ell$ physically corresponds to an impact parameter that the photons can not reach. Contrary, for orbits of second kind, with $0<E<E_u$ ($b_u<b
<\infty$), the photons can have an impact parameter $\ell$ by describing the Hippopede geodesic, with a return point $r_0$, see Fig. \ref{f3}.
Note that for $Q=0$, $r_u(Q=0)=2\sqrt{2}M$, and it is greater than $r_u$ for RNAdS; however, the maximum value of the potential $E_u(Q=0)$ is smaller than $E_u$ for RNAdS. Also, the $r_0$ value is the same for both spacetimes.

Next, based on the impact parameter values 
and Fig. \ref{f3},
we present a brief qualitative description of the allowed
angular motions for photons in RNAdS.

\begin{itemize}
	\item \emph{Capture zone}:
	If $0<b <b_{u}$, photons fall inexorably
	to the  horizon $r_+$, or escape to infinity, depending on initial conditions,
	and its cross section,
	$\sigma$, in these geometry is \cite{wald}
	\begin{equation}\label{mr51}
	\sigma=\pi\,b_u^2={ \pi\,r_u^2\over f(r_u)}.
	\end{equation}
	\item \emph{Critical trajectories}:
	If $b=b_{u}$, photons can stay in one of the unstable
	inner circular orbits of radius  $r_{u}$.
	Therefore, the photons that arrive from the initial distance
	$r_i$ ($r_+ < r_i< r_u$, or $r_u< r_i<\infty$)
	can asymptotically fall to a circle of radius $r_{u}$.
	The affine period in such orbit is
	\begin{equation}\label{p1}
	T_{\lambda}=\frac{2\pi\,r_u^2}{L}\,,
	\end{equation}
	and the coordinate period is
	\begin{equation}\label{p2}
	T_t=2\pi\,b_u=\frac{2\pi\,r_u}{\sqrt{f(r_u)}}\,.
	\end{equation}
	\item \emph{Deflection zone}. If $b_{u} <b <b_{\ell}$, this zone presents orbits of first and second kind. The orbits of first kind are allowed in the interval $r_D\leq r<\infty$, where the 
	photons can come from a finite distance or from an infinity distance until they reach the distance $r=r_{D}$ (which is solution of the equation $V(r_D)=E$), and then the photons are deflected. Note that photons with $b\geq b_{\ell}=\ell$ are not allowed in this zone. The orbits of second kind are allowed in the interval $r_+<r \leq r_F$, where the photons come from a distance greater than the event horizon, then they reach the distance $r_F$  (which is solution of the equation $V(r_F)=E$) and then they plunge into the horizon. 
	\\

\item \emph{ Second kind and Hippopede geodesic  }. If $ b_u<b<\infty$, the return point is in the range  $r_+<r<r_u$, and then the photons plunge into the horizon. However, when $b=b_{\ell}$ a special geodesic can be obtain, known as  Hippopede of Proclus.

\end{itemize}

\begin{figure}[!h]
	\begin{center}
		\includegraphics[width=90mm]{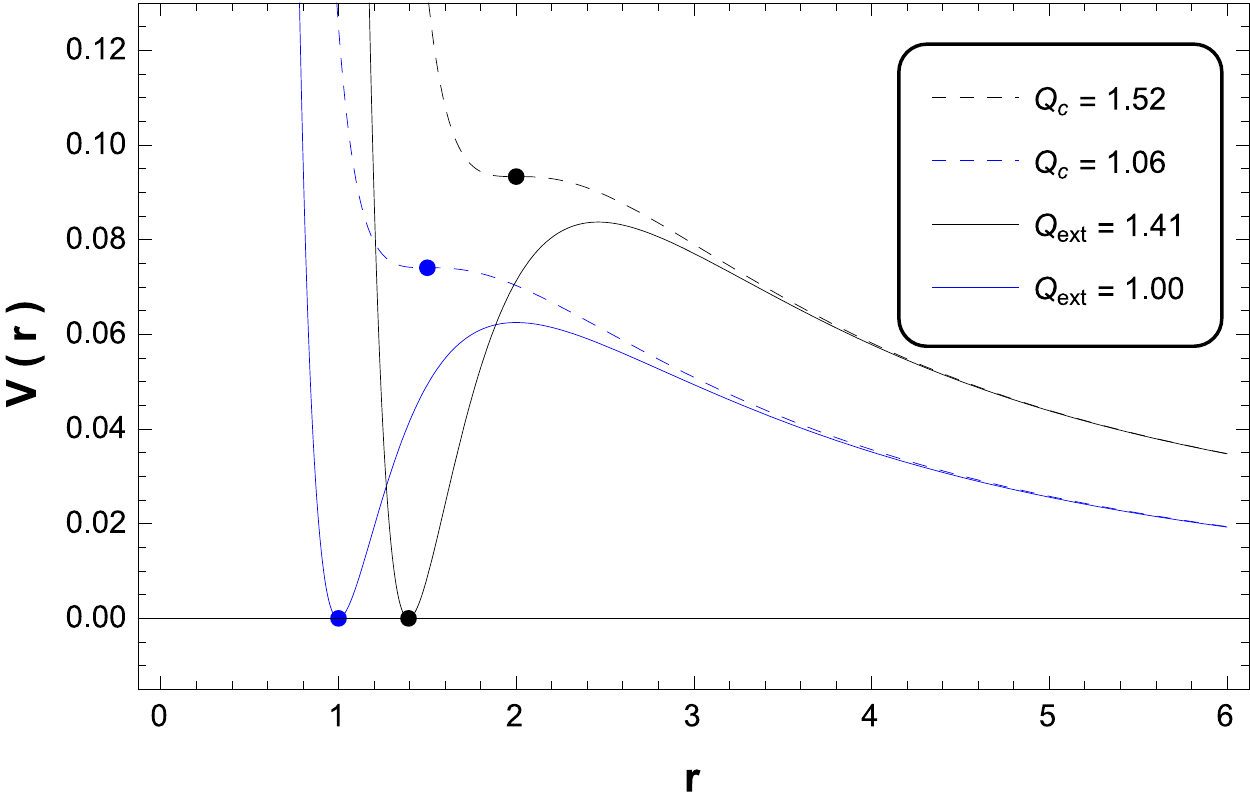}
	\end{center}
	\caption{Plot of the effective potential of photons as a function of $r$ for different values of the charge $Q$.
		Here we have used the values $L=1$, and $M=1$. Blue line corresponds to four-dimensional RN spacetime $Q_{ext}=1.0$  $r_+=1$, $Q_c=1.06$, and the radius of the last unstable circular orbit is $r_u=1.5$. Black line corresponds to five-dimensional RN AdS spacetime with $\ell=10$, $Q_{ext}=1.41$, $r_+=1.39$, $Q_c=1.52$, and the radius of the last unstable circular orbit is $r_u=2$.}
	\label{criticalvalue}
\end{figure}

On the other hand, it was argued that an introduction of a negative tidal charge in four-dimensional Reissner–Nordström black holes  can describe black hole solutions in theories with extra dimensions in Ref. \cite{Kuniyal:2017dnu}. Also, by considering a naked singularity, i.e, $q=Q^2/M^2>1$ it was shown the existence of a critical value of $q=q_c=9/8$ for shadow existence; thereby, for $q\leq 9/8$  the Reissner–Nordström spacetimes have shadows and the radius of the last unstable circular orbit is $r_{u}= 3M/2$, while that  for $q>9/8$ the  shadows  do  not  exist. Interestingly, at the same critical value the quasinormal modes for the scattering exhibit a different behavior \cite{Chirenti:2012fr}, as well as, it is responsible for the existence of circular orbits of neutral test particles \cite{Pugliese:2010ps}. The critical charge $q_c$, arises from the last unstable circular orbits considering a naked singularity; thus, for five-dimensional RNAdS spacetimes, from Eq. (\ref{max1}), one can deduce the critical value of the charge where occurs the last unstable circular orbit given by  $Q_c^2= {4M^2\over \sqrt{3}}$, so $q_c= {4\over \sqrt{3}}$, and the radius of the last unstable circular orbit is $r_{u}=2M$.  
Also for $Q > Q_{ext}$, where 
\begin{equation}
Q_{ext}=\left[\frac{2 \ell}{27}\left[\left(\ell ^2+12 M^2\right)^{3/2}-\ell \left(\ell ^2+18 M^2\right)\right]\right]^{1/4}\,,
\end{equation}
the spacetime describes a naked singularity, where $Q_{ext}$ was obtained using Eq. (\ref{equ}). 
It is worth noticing that  $q_c$ does not depend on the cosmological constant. However, the value of the charge for which the spacetime describes a naked singularity depends on the value of the cosmological constant. In Fig. \ref{criticalvalue}, we show the behaviour of the effective potential as a function of $r$, where the points indicate the radius of the last unstable circular orbit for the critical charge $q_c$ and the event horizon for the extremal charge $Q_{ext}$. We can observe that the critical charge $q_c$, $Q_{ext}$, and the radius of the last unstable circular orbit increase when the spacetime is the five-dimensional RNAdS instead of the four-dimensional Reissner–Nordström.  

\newpage

\subsection{Bending of light}

Now, in order to obtain the bending of light we consider
Eq. (\ref{eqmot}), which can be written as 
\begin{equation}\label{t7}
\left(r\frac{dr}{d\psi}\right)^2={r^6-\mathcal{B}^2r^4+4M^2\mathcal{B}^2r^2-Q^4\mathcal{B}^2\over \mathcal{B}^2}
={\mathcal{P}(r)\over \mathcal{B}^2}\,.
\end{equation}
So, in order to obtain the return points, we solve the $\mathcal{P}(r)=0$. Thus, by performing 
the change of variable $y=r^2+\mathcal{B}^2 /3$, $\mathcal{P}(y)=y^3-\tilde{\alpha} y-\tilde{\beta}$,
where
\begin{equation}
\tilde{	\alpha}=\mathcal{B}^2\left( \frac{\mathcal{B}^2}{3}-4M^2\right) 
	, \,\,\,\tilde{ \beta}=\mathcal{B}^2\left( Q^4-\frac{4M^2\mathcal{B}^2}{3}+\frac{2\,\mathcal{B}^4}{27}\right)\,,
\end{equation}
and the deflection distance $r_{D}$ is given by
\begin{eqnarray}\label{sol1}
	r_{D}&=&\sqrt{\chi_0\cos \chi_1+{\mathcal{B}^2\over3}}, \\\label{sol2}
\end{eqnarray}
and the return point $r_F$ is
\begin{eqnarray}
	\label{sol2}
	r_{F}&=&\sqrt{{\chi_0\over2}\left(\sqrt{3}\sin \chi_1- \cos \chi_1\right) +{\mathcal{B}^2\over3}}\,,
\end{eqnarray}
where $\chi_0=2 \sqrt{\tilde{\alpha}/3}$ and $\chi_1=\frac{1}{3} \arccos \left( \frac{3 \tilde{\beta}}{2 }\sqrt{\frac{3}{\tilde{\alpha}^{3}}} \right)$.

\begin{figure}[!h]
	\begin{center}
		\includegraphics[width=25mm]{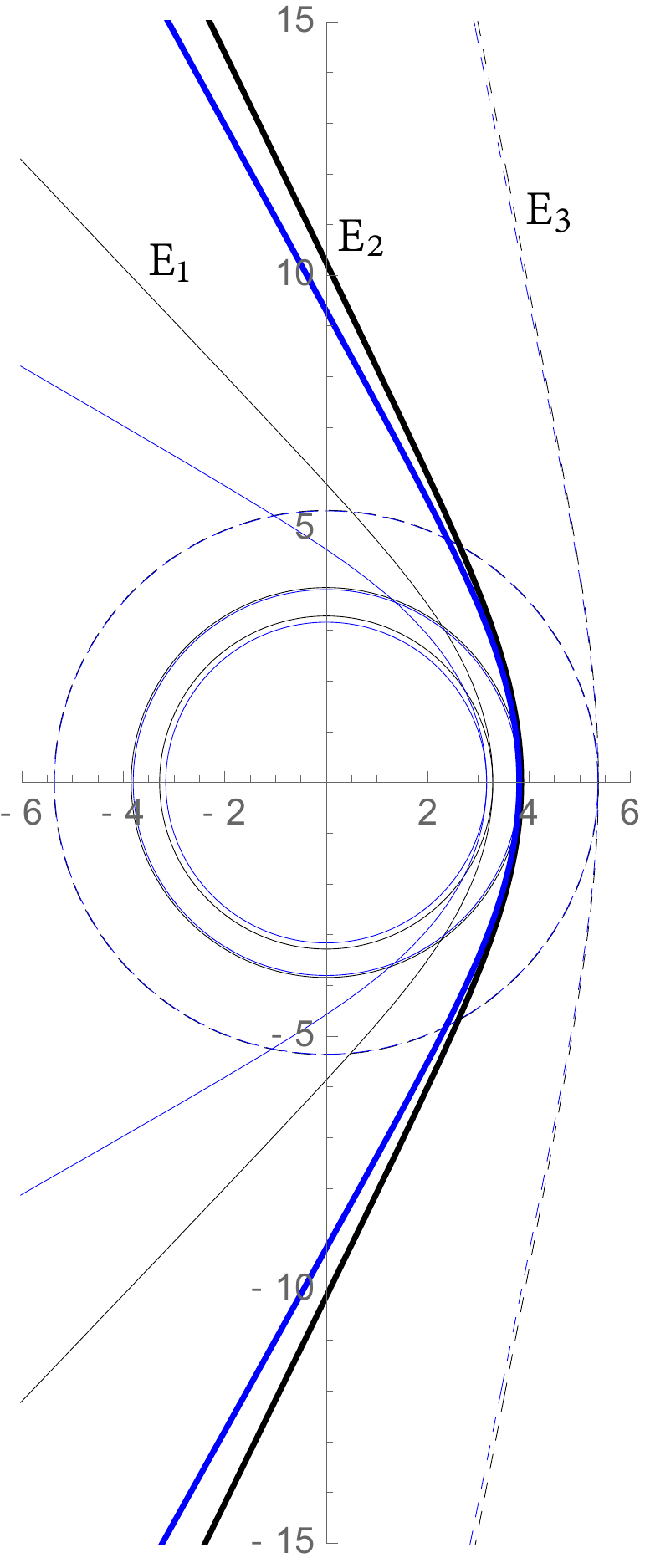}
	\end{center}
	\caption{Polar plot for deflection of light with $\ell=10$, and $L=1$. Thin line $E_1=0.07$, thick line $E_2=0.06$, and dashed line $E_3=0.04$. Black lines for $Q=1.20$ and blue lines for $Q=0$.}
	\label{DFoton}
\end{figure}

Then, after a brief manipulation, and performing the change of variable $r=\mathcal{B}\sqrt{4x+1/3}$ it is possible
to integrate Eq. (\ref{t7}), given the following expression 
\begin{equation}
\label{cuadrat}
\psi=\int^{x}_{x_D}{dx\over 2 \sqrt{4x^3-g_2 x-g_3}}\,,
\end{equation}
where the invariants are given by 
\begin{eqnarray}
\label{inv1} g_2&=&{1\over 12}-{M^2\over \mathcal{B}^2},\\ \label{inv2} g_3&=&{1\over 16}\left(\frac{2}{27}-\frac{4 M^2}{3\mathcal{B}^2}+\frac{Q^4}{\mathcal{B}^4}\right)\,.
\end{eqnarray}
Therefore, by integrating the Eq. (\ref{cuadrat}) and then solving for $r$ leads to
\begin{equation}
r(\psi)=\mathcal{B}\sqrt{4\wp(2\psi+\omega_D)+1/3}\,,
\label{mr.6}
\end{equation}
where $\omega_D=\wp^{-1}(r_D^2/4\mathcal{B}^2-1/12)$. In Fig. \ref{DFoton} we show the behaviour of the bending of light.  We can observe that the deflection angle is greater, when the black hole is uncharged. On the other hand, note that the above equations are straightforward obtained for 
five-dimensional Schwarzschild anti-de Sitter spacetime.

\subsubsection{The deflection angle}

\begin{figure}[!h]
	\begin{center}
		\includegraphics[width=8cm]{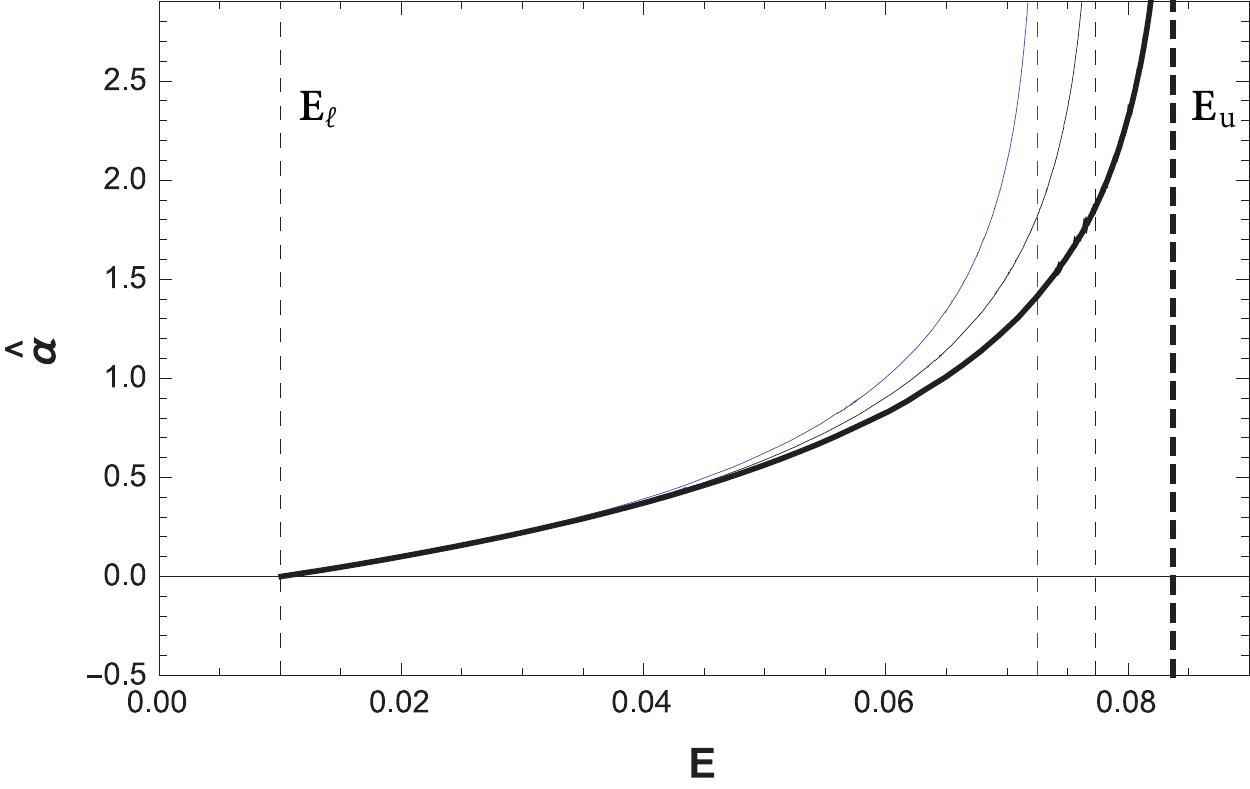}
	\end{center}
	\caption{The behaviour of the deflection angle $\hat{\alpha}$ in terms of $E$, demonstrated for $L=1$, $M=1$ and $\ell=10$. Black-thin line for $Q=1.2$, black-thick line $Q=Q_{ext}=1.40$, and  blue line for $Q=0$. As it is expected, the deflection angle reaches its limit as $E$ tends to $E_u$ which for $Q=1.20$ is around $0.077$, for $Q=Q_{ext}$ is $0.084$ and it is $0.073$ for $Q=0$.}
	\label{fig:anguloalfa1}
\end{figure}

It is known that photons can escape to infinity during a scattering process. So, by considering  $r(\psi)|_{\psi=0} = r_D$ is the shortest distance to the black hole 
at which the deflection happens, and assuming that the incident photons are coming from infinity and escape to infinity, so $r(\psi)|_{\psi=\psi_{\infty}} = \infty$. Now, by using Eq. (\ref{mr.6}) we obtain that $2\psi_{\infty}=-\omega_D$, and the deflection angle $\hat{\alpha} = 2\psi_\infty-\pi$, is given by
\begin{equation}\label{mr.8}
\hat{\alpha} =- \wp^{-1}(r_D^2/4\mathcal{B}^2-1/12)-\pi\,.
\end{equation}
The evolution of the deflection angle has been plotted in Fig.~\ref{fig:anguloalfa1} which has an asymptotic behavior as $E\rightarrow E_u$. We can observe that the deflection angle takes an infinity value when the $E=E_u$, such $E_u$ increases when the charge of the black hole increases.

\subsection{Second kind trajectories and Hippopede  geodesic}

\begin{figure}[!h]
	\begin{center}
		\includegraphics[width=5cm]{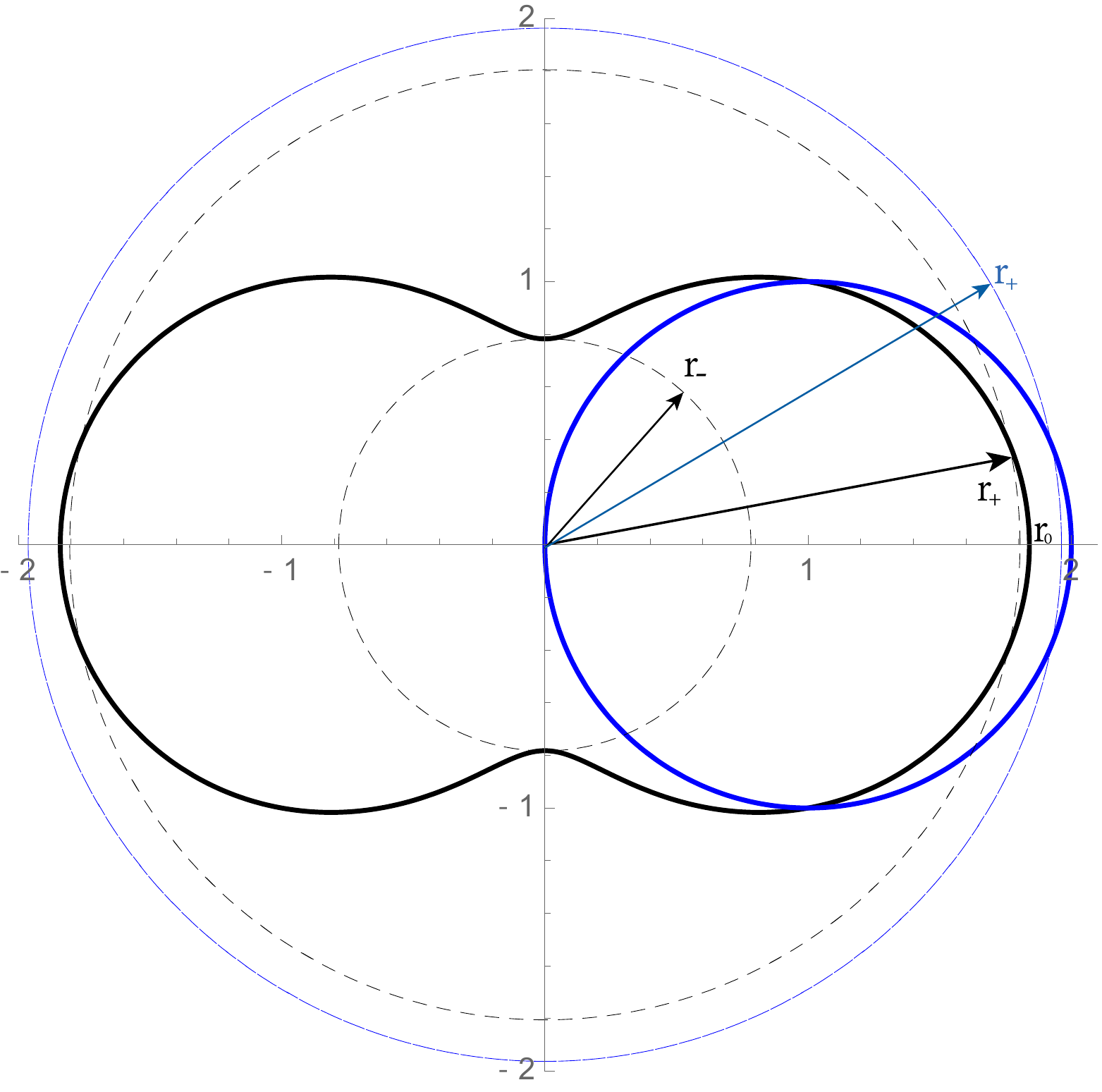}
	\end{center}
	\caption{The hippopede geodesic (black line), with $E_{\ell}=0.01$, $Q=1.20$, and $\ell=10$, dashed black lines correspond to the horizons. The circumference geodesic (Blue lines), with $E_{\ell}=0.01$, $Q=0$, and $\ell=10$.}
	\label{fig:anguloalfa}
\end{figure}

The spacetime allows second kind trajectories, when $ b_u<b<\infty$, where the return point is in the range $r_+<r<r_u$, and then the photons plunge into the horizon. However, a special geodesic can be obtained when
the anomalous impact parameter $\mathcal{B} \rightarrow \infty$ ($b= \ell$). 
In this case, the radial coordinate is restricted to $r_+ < r < r_0$, and the equation of motion (\ref{eqmot}) can be written as
\begin{equation}
\label{hipo1}
\psi=-\int^{r}_{r_0}{r\,dr\over  \sqrt{-r^4+4M^2r^2-Q^4}}\,,
\end{equation}
and the return points are:
\begin{eqnarray}
\label{r0} r_0&=&2M\cos\left[ {1\over 2}\sin^{-1}\left({Q^2\over 2M^2} \right) \right]\,,\\ \label{rho0} \rho_0&=&2M\sin\left[ {1\over 2}\sin^{-1}\left({Q^2\over 2M^2} \right) \right]\,.
\end{eqnarray}
Thus, it is straightforward to find the solution of Eq. (\ref{hipo1}), which is given by
\begin{equation}
r(\psi)=\sqrt{2M^2+\sqrt{4M^4-Q^4}\,\cos[2\psi]}\,,
\label{hipo2}
\end{equation}
which represents to the
Hippopede of Proclus geodesic (see Fig. \ref{fig:anguloalfa}) \cite{lawrence}. This trajectory is a new type of orbits in five-dimensional RNAdS, and it does not depend on the value of the cosmological constant. It is worth mentioning that the analogue geodesic in four-dimensional RNAdS corresponds to the Lima\c{c}on of Pascal \cite{Villanueva:2013zta}. Also, when the spacetime is the five-dimensional Schwarzschild anti-de Sitter this geodesic is given by $r=2M\cos[\psi]$, which describes a circumference with radius $M$, that is analogue to the cardioid geodesics found in four-dimensional Schwarzschild anti-de Sitter \cite{Cruz:2004ts}.

\subsection{Critical trajectories  and  capture zone}
In the case of $b=b_u$, the particles can be confined on unstable circular orbits of the radius $r_u$. This kind of motion is indeed ramified into two cases; critical trajectories of the first kind (CFK) in which the particles come from a distant position $r_i$ to $r_u$ and those of the second kind (CSK) where the particles start from an initial point $d_i$ at the vicinity of $r_i$ and then tend to this radius by spiraling. We obtain the following equations of motion for the aforementioned trajectories:

\begin{equation}
r(\psi)=\left[ \rho_u^2+(r_u^2-\rho_u^2)\left({1+C\,e^{\kappa\,\psi}\over 1-C\,e^{\kappa\,\psi}} \right) ^2\right] ^{1/2}\,,
\label{criticas}
\end{equation}
where
\begin{eqnarray}
\label{rh0} \rho_u&=&\sqrt{\mathcal{B}_u^2-2r_u^2}\,,\\ \label{C} C&=&\left| {\sqrt{r_i^2-\rho_u^2}-\sqrt{r_u^2-\rho_u^2}\over \sqrt{r_i^2-\rho_u^2}+\sqrt{r_u^2-\rho_u^2}}\right|\,, \\
\label{kappa} \kappa&=&{2\sqrt{r_u^2-\rho_u^2}\over \mathcal{B}_u}\,.
\end{eqnarray}

\begin{figure}[!h]
	\begin{center}
		\includegraphics[width=8.0cm]{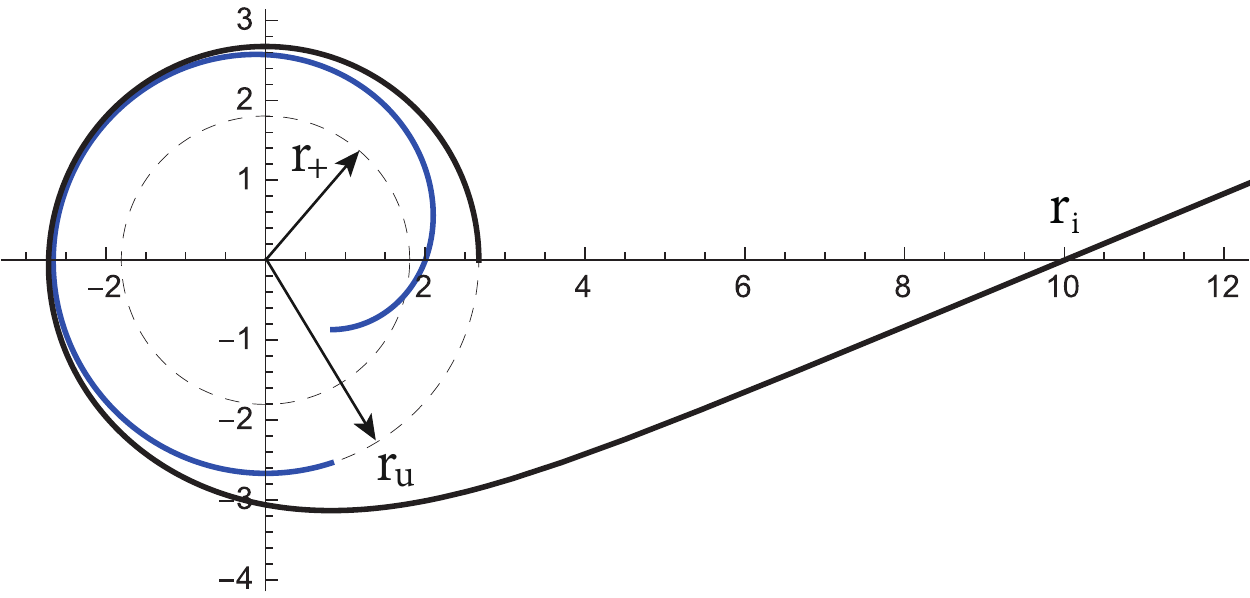}
	\end{center}
	\caption{The critical trajectories $r(\psi)$ plotted for $Q=1.2$, $L=1.0$ $\ell=10$, with $E_u \approx 0.077 $, $r_u \approx 2.67$, and $r_i=10$. Black line for CFK and blue line for CSK trajectories.}
	\label{fig:critical}
\end{figure}

In Fig. \ref{fig:critical}, we show the behaviour of the CFK and CSK trajectories, given by Eq. (\ref{criticas}). Note that for the second kind trajectories,  $r_i$ must be replaced by $d_i$ in the constant $C$ (\ref{C}). On the other hand, for photons with an impact parameter less than the critical one ($b_u$), which are in the capture zone, they can plunge into the horizon or escape to infinity, with a cross section given by Eq. (\ref{mr51}).

\subsection{Shapiro time delay}

An interesting relativistic effect in the propagation of light rays is the apparent delay in the time of propagation for a light signal passing near the Sun, which is a relevant correction for astronomic observations, and is called the Shapiro time delay. The time delay of Radar Echoes corresponds to the determination of the time delay of radar signals which are transmitted from the Earth through a region near the Sun to another planet or  spacecraft and then reflected back to the Earth. The time interval between emission and return of a pulse as measured by a clock on the Earth is

\begin{equation}
t_{12}=2\, t(r_1,r_D)+2\, t(r_2,r_D)~,
\end{equation}
where $r_D$ as closest approach to the Sun. Now, in order to calculate the time delay we use Eq. (\ref{w.12}), 
and by considering that
$dr/dt$ vanishes, thereby
$\frac{E^2}{L^2}=\frac{f(r_D)}{r_D^2}$.
Thus, the coordinate time that the light requires to go from $r_D$ to $r$ is given by
\begin{equation}
t(r,r_D)=\int_{r_D}^r \frac{dr}{f(r)\sqrt{1-\frac{r_D^2}{f(r_D)}\frac{f(r)}{r^2}}}~.
\end{equation}
So, at first order correction we obtain
\begin{equation}
 t(r, r_D)=\sqrt{r^2-r_D^2}+t_M(r)+t_Q(r)+t_{\ell}(r)\,,
\end{equation}
where
\begin{eqnarray}
t_M(r)&=&\frac{6M^2}{r_D}\sec^{-1}\left({r\over r_D}\right)\,,\\
t_Q(r)&=&-\frac{Q^4}{4r_D^3}\left[5\sec^{-1}\left({r\over r_D}\right)+{3\,r_D\sqrt{r^2-r_D^2}\over r^2}\right]\,,\\
t_{\ell}(r)&=&-{\sqrt{r^2-r_D^2}\over 3 \ell^2}\left(r^2+{r^2_D\over 2}\right)\,.
\end{eqnarray}
Therefore, for the circuit from point 1 to point 2 and back the delay in the coordinate time is
\begin{equation}
\Delta t := 2\left[t(r_1, r_D)+t(r_2,r_D)-\sqrt{r_1^2-r_D^2}-\sqrt{r_2^2-r_D^2}\right]~,
\end{equation}
where
\begin{eqnarray}
\notag
\Delta t &=&2\left[t_M(r_1)+t_M(r_2)+t_Q(r_1)+t_Q(r_2)\right]\\
&& +2\left[t_{\ell}(r_1)+t_{\ell}(r_2)\right]~.
\end{eqnarray}

Now, for a round trip in the solar system, we have ($r_D <<r_1,r_2$)
\begin{eqnarray}
\label{deltat}
\notag
\Delta t &\approx& \left[ \frac{12M^2}{r_D}-\frac{5\,Q^4}{2\,r_D^3}\right]\left[\sec^{-1}\left({r_1\over r_D}\right)+\sec^{-1}\left({r_2\over r_D}\right)\right]\\
&& -\frac{2}{3\,\ell^2}\left(r_1^3+r_2^3\right)~.
\end{eqnarray}

 Note that the classical result of GR; that is, $\Delta t_{GR}=4M_{\odot}\left[ 1+ ln\left(\frac{4r_1r_2}{r_D^2}\right)\right]$. For a round trip from the Earth to Mars and back, we get (for $r_D \ll r_1 , r_2$ ), where $r_1 \approx r_2=2.25\times 10^{8}Km$  is the average distance Earth-Mars. Considering $r_D$, as closest approach to the Sun, like the radius  of the Sun ($R_{\odot} \approx 696000\,Km$) plus the solar corona ($  \sim 10^{6}Km$), $r_D \approx1.696\times 10^{6}Km$, then,  the time delay is $\Delta t_{GR} \approx 240\, \mu\,s$. On the other hand, if we consider the limit ${M}\rightarrow M_{\odot}$, $Q=0$, and $\Lambda=0$, in Eq. (\ref{deltat}), we obtain $\Delta t \approx 161 ns$. It is worth to mention that this value is more near to the value measure in the Viking mission, where the error in the time measurement of a circuit was only about $10\, ns$ \cite{Straumann}.

\section{Final remarks}
\label{conclusion}

We considered the motion of photons in the background of five-dimensional RNAdS black holes, and we established the null structure geodesic. This spacetime is described by one Cauchy horizon and an event horizon. Concerning to the radial motion, we showed that as seen by a system external to photons, they will fall asymptotically to the event horizon. On the other hand, this external observer will see that photons arrive in a finite coordinate time to spatial infinite. Concerning to the angular motion, we found analytically orbit of first and second kind; and critical orbit. Interestingly,  for second kind trajectory, we found that the motion of photons follows the hippopede of Proclus geodesic when the parameter of impact $b$ takes the value $b=\ell$, and it does not depend on the value of the cosmological constant, being the Lima\c{c}on of Pascal their analogue geodesic in four-dimensional RNAdS. On the other hand, we studied some observational test such as the bending of light, which have a similar behaviour that four-dimensional RNAdS, and the Shapiro time delay effect, where our results show that $\Delta t \approx 161 ns$ while that for GR $\Delta t_{GR} \approx 240\, \mu\,s$.

Also, by comparing five-dimensional RNAdS black holes with four-dimensional RNAdS black holes, for the five-dimensional spacetime  there is not an additional parameters apart of the dimension added, contrary to five-dimensional Myers-Perry black hole spacetime, where the metric describes a spacetime with two spin parameters
which could explain the differences with respect to four-dimensional Kerr black hole spacetime due to the presence of two spin parameters in higher dimension.  For five-dimensional RNAdS black holes the event horizon is not the same due to the change in the lapse function, which could explain the differences between four and five-dimensional spacetimes. However, the effect of a additional dimension could be the existence of the hippopede of Proclus geodesic found here, versus its analogue geodesic in four-dimensional RNAdS black hole, i.e, the Lima\c{c}on of Pascal \cite{Villanueva:2013zta}, both trajectories of second kind.\\

{\bf{Acknowledgments}}\\

We thank the referee for his/her careful review of the manuscript and his/her valuable comments and suggestions. Y.V. acknowledge support by the Direcci\'on de Investigaci\'on y Desarrollo de la Universidad de La Serena, Grant No. PR18142. J.R.V. was partially supported by Centro de Astrof\'isica de Valpara\'iso.

\end{document}